\documentclass[prl,twocolumn,showpacs,preprintnumbers,superscriptaddress]{revtex4-1}
\usepackage{graphicx}

\usepackage{dcolumn}
\usepackage{color}

\newcommand{\beq}{\begin{equation}}
\newcommand{\eeq}{\end{equation}}
\newcommand{\beqa}{\begin{eqnarray}}
\newcommand{\eeqa}{\end{eqnarray}}

\begin{document}

\title{Critical properties of the half-filled Hubbard model in three dimensions}
\author{G.~Rohringer}
\affiliation{Institute for Solid State Physics, Vienna University of Technology 1040
Vienna, Austria}
\author{A.~Toschi}
\affiliation{Institute for Solid State Physics, Vienna University of Technology 1040
Vienna, Austria}
\author{A.~Katanin}
\affiliation{Institute of Metal Physics, Ekaterinburg, Russia}
\author{K.~Held}
\affiliation{Institute for Solid State Physics, Vienna University of Technology 1040
Vienna, Austria}
\pacs{71.10.Fd, 71.27.+a}

\begin{abstract}
By means of the dynamical vertex approximation (D$\Gamma$A) we include spatial
correlations on all length scales beyond the dynamical mean field theory
(DMFT) for the half-filled  Hubbard model in three dimensions. The most relevant changes
due to non-local fluctuations are: (i) a deviation from the mean-field critical behavior
with the same critical exponents as for the three dimensional Heisenberg (anti)-ferromagnet
and (ii) a sizable reduction of the N\'eel temperature ($T_N$) by $\sim 30 \%$ for the onset of antiferromagnetic order.
Finally, we give a quantitative estimate of the deviation of the spectra between D$\Gamma$A and DMFT in different regions of the phase-diagram.
\end{abstract}

\date{\today}
\maketitle

Almost 50 years after the invention of the Hubbard model \cite{Hubbard63a} and despite modern petaflop supercomputers, a precise analysis of the criticality of this most basic model for electronic
correlations has not been achieved so far, at least not in three
dimensions. Dynamical mean field theory (DMFT) \cite{DMFT,DMFT2,DMFTreview}
was a big step forward to calculate
the three dimensional Hubbard model since the major contribution of
electronic correlations, i.e, the local one, is well captured within
this theory. Local correlations give rise to quasiparticle
renormalization, the Mott-Hubbard transition, magnetism, and even more
subtle issues such as  kinks in
purely electronic models \cite{kinks}. However, non-local spatial
correlations are also naturally generated by a purely local Hubbard
interaction, and, as it is well known, they become of essential importance
in the vicinity of second-order phase transitions. As these correlations are neglected in DMFT, this scheme provides only for a conventional mean-field (MF) description of the critical properties.

To overcome this shortcoming cluster extensions to DMFT such as the
dynamical cluster approximation (DCA) and cluster-DMFT have been proposed
\cite{Maier04}. In these approaches spatial
correlations beyond DMFT are taken into account, however only within the
range of the cluster size; and due to computational limitations the actual
size of  $d\!=\!3$-clusters is restricted to about 100 sites. Hence, short-range correlations are included by these approaches, whereas long-range ones are not (e.g. for spacings larger than 5 lattice sites in $d=3$). Nonetheless, Kent \emph{et al.}
\cite{Kent05a} were able to extrapolate the cluster size of so-called Betts
clusters to infinity, albeit assuming from the beginning the critical
exponents to be those of the Heisenberg model. This way they extrapolated
the  N\'eel temperature of the paramagnetic-to-antiferromagnetic phase
transition which was found in agreement with earlier lattice quantum Monte
Carlo (QMC) results by Staudt \emph{et al.} \cite{Staudt00}.

As an alternative to cluster extensions and, in particular, to
include long-range correlations on an equal footing, more recently diagrammatic
expansions of DMFT have been proposed: (i) the DMFT plus spin-fermion model \cite{sigma_k}, (ii) the dynamical vertex approximation (D$\Gamma $A) \cite{DGA1,Kusunose06a,Slezak06a} which approximates the fully irreducible $n$-particle vertex to be local \cite{DGA1} or that of a DCA cluster \cite{Slezak06a}; and (iii) the dual fermion
approach \cite{DualFermion}. As for phase transitions,  D$\Gamma$A  with Moriyasque corrections \cite{DGA2} fulfills - in contrast with dual fermion calculations of \cite{DualFermion2}- the Mermin and Wagner theorem in two-dimensions  and, as we will discuss in the following, corrects the MF behavior for the critical exponents in three dimensions.

In this paper, we apply the aforementioned approximation of the D$\Gamma $A scheme (with Moriyasque corrections \cite{DGA2}) for studying the phase-diagram of the
three dimensional Hubbard model at half-filling. In particular, we (i) calculate the critical exponents, (ii) determine the phase diagram with $T_N$ substantially reduced compared to the DMFT one, and  (iii) define the region where non-local correlations become too strong so that DMFT is not applicable anymore.

%

 We consider  the Hubbard model on a cubic lattice
\begin{equation}
H=-t\sum_{\langle ij\rangle \sigma }c_{i\sigma }^{\dagger }c_{j\sigma
}+U\sum_{i}n_{i\uparrow }n_{i\downarrow }  \label{H}
\end{equation}%
where $t$ denotes the hopping amplitude between nearest-neighbors, $U$ the
Coulomb interaction, and $c_{i\sigma }^{\dagger }$($c_{i\sigma }$) creates
(annihilates) an electron with spin $\sigma $ on site $i$; $n_{i\sigma
}\!=\!c_{i\sigma }^{\dagger }c_{i\sigma }$. In the following, we restrict
ourselves to the paramagnetic phase with $n=1$ electron/site at a finite
temperature $T$. For the sake of clarity, and in accordance with previous
publications, we will define hereafter our energies in terms of a typical
energy scale $D=2 \sqrt{6}\, t$ \cite{notaDOS}.

The D$\Gamma$A approach to the model (\ref{H}) was derived in Refs.
\onlinecite{DGA1,DGA2}. The dynamic non-uniform susceptibility reads
\begin{equation}
\chi _{\mathbf{q}\omega }^{s(c)}=[(\phi _{\mathbf{q,}\omega
}^{s(c)})^{-1}\mp U+\lambda _{s(c)}]^{-1}
\end{equation}
with $\phi _{\mathbf{q},\omega }^{s(c)}\!=\!\sum\limits_{\nu \nu ^{\prime }}\Phi
_{s(c),\mathbf{q}}^{\nu \nu ^{\prime }\omega } \!=\! \sum\limits_{\nu \nu ^{\prime }}  [(\chi _{0\mathbf{q}\omega }^{\nu ^{\prime }})^{-1}\delta _{\nu}^{ \nu ^{^{\prime }}}\!-\!\Gamma _{s(c)\text{ir}}^{\nu \nu ^{\prime }\omega }\!\pm\! U]^{-1},
$ $\chi _{0\mathbf{q}\omega }^{\nu ^{\prime }}=-T\sum_{\mathbf{k}}G_{\mathbf{k}
,\nu ^{\prime }}G_{\mathbf{k}+\mathbf{q},\nu ^{\prime }+\omega }$
(particle-hole bubble), $G_{\mathbf{k},\nu }=[i\nu -\epsilon _{\mathbf{k}}+\mu
-\Sigma _{\text{loc}}(\nu )]^{-1}$ (Green function) and $\Sigma _{
\text{loc}}(\nu )$ (local self-energy). The vertex $\Gamma _{s(c),\text{ir}
}^{\nu \nu ^{\prime }\omega }$ is determined from the solution of the
single-impurity problem\cite{DGA1}.
 In fact, the complete inclusion of non-local corrections in the irreducible vertices in all channels
can be achieved only via the fully self-consistent D$\Gamma $A equations.
However, as discussed in Ref.\ \cite{DGA2}, when considering a situation
where no competition between different instabilities occurs,
a restriction to one specific
channel and the evaluation of the self-consistency effect via the
corresponding Moriyasque correction $\lambda _{s(c)}$ is possible \cite{DGA2}. In the half-filled case we neglect non-local particle-particle fluctuations
since this channel is strongly suppressed by the repulsive interaction. Furthermore, at half filling,
charge excitations are generically expected to be irrelevant for the critical behavior as well.
Indeed we find $\chi^c_{{\mathbf q} \omega},\chi^{pp}_{{\mathbf q} \omega}\ll  \chi^s_{{\mathbf q} \omega}$ ($\chi^{pp}_{{\mathbf q} \omega}$ is the particle-particle susceptibility), 
hence we
neglect non-local particle-particle contributions as well as  $\lambda _{c}$ and determine $\lambda _{s}$ from the exact sum rule  (which also holds for DMFT)
$
-\int_{-\infty }^{\infty }\frac{d\nu }{\pi }\mbox{Im}\Sigma _{\mathbf{k},\nu
}=U^{2}n(1-n/2)/2,  \label{As1}
$
where the non-local self-energy is given by
\begin{eqnarray}
\Sigma&&_{\mathbf{k},\nu }=\frac{1}{2}{Un}+\frac{1}{2}TU\sum\limits_{%
\omega ,\mathbf{q}}\left[ 3\gamma _{s,\mathbf{q}}^{\nu \omega }-\gamma _{c,%
\mathbf{q}}^{\nu \omega }-2+3U\gamma _{s,\mathbf{q}}^{\nu \omega }\chi _{\mathbf{q}\omega
}^{s}\right.   \nonumber \\
&&\;\;+U\gamma _{c,\mathbf{q}}^{\nu \omega }\chi _{\mathbf{q}\omega }^{c}
+\sum\limits_{\nu ^{\prime }}\left. \chi _{0\mathbf{q}\omega }^{\nu
^{\prime }}(\Gamma _{c,\text{loc}}^{\nu \nu ^{\prime }\omega }-\Gamma _{s,%
\text{loc}}^{\nu \nu ^{\prime }\omega })\right] G_{\mathbf{k+q},\nu +\omega } 
\label{SE0}
\end{eqnarray}
with $ \gamma _{s(c),\mathbf{q}}^{\nu \omega }=(\chi _{0\mathbf{q}\omega }^{\nu
})^{-1}\sum\limits_{\nu ^{\prime }}\Phi _{s(c),\mathbf{q}}^{\nu \nu ^{\prime
}\omega },  \label{gamma} $
and $\Gamma _{s(c),\text{loc}}^{\nu \nu ^{\prime }\omega }$ is the reducible
local spin (charge) vertex, determined from the single-impurity problem.


Starting point of our investigation of the critical properties of the
antiferromagnetic (AF) instability is the corresponding (divergent) spin susceptibility
\begin{equation}
\chi _{AF}=\chi _{\mathbf{Q,}0}^{s}=\int\limits_{0}^{\beta }d\tau \langle
S_{z,\mathbf{Q}}(\tau )S_{z,-\mathbf{Q}}(0)\rangle
\end{equation}%
with $\mathbf{Q}=(\pi ,\pi ,\pi )$. While the D$\Gamma $A with Moriyasque
corrections well reproduces the textbook Mermin and Wagner results for the Hubbard model in  $d\! = \! 2$ yielding finite, but exponentially large susceptibility at finite $T$ \cite{DGA2}, the situation in $d\! = \! 3$ is even more
intriguing, since the AF-phase remains stable in a broad region at finite $T$, allowing for a direct study of the critical properties.

\begin{figure}[tb]
\begin{center}
\includegraphics[width=8cm]{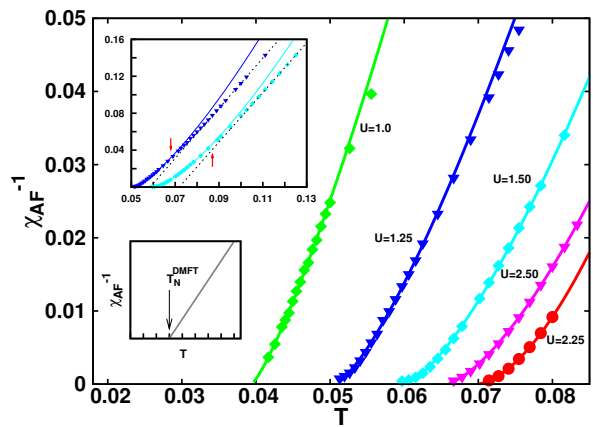} 
\end{center}
\par
\vspace{-.5cm}
\caption{(Color online) Inverse  AF spin susceptibility as a function of $T$ for different $U$ values. Lower inset: Inverse DMFT susceptibility with a MF ($\gamma=1$: linear behavior) critical exponent. Upper inset: larger $T$ interval. }
\label{AFsusc}
\end{figure}

Of particular interest is the analysis of the evolution of the critical
region as a function of the Coulomb repulsion. In Fig.\ \ref{AFsusc}, we
show the inverse susceptibility $\chi _{AF}^{-1}$ as a function of $T$ for
different $U$ values. The vanishing of $\chi _{AF}^{-1}\propto (T-T_{N})^{\gamma }$ marks the onset of
the AF long-range order, defining the corresponding $T_{N}$ for a given $U$%
. More important is, however, the examination of the critical behavior:
While in a MF (or DMFT) approach $\chi _{AF}^{-1}$ is vanishing linearly
close to $T_{N}$ in accordance with the MF (Gaussian) critical exponent $\gamma  =  1$ (see lower inset of Fig.\ \ref{AFsusc}), D$\Gamma $A data clearly show a bending in the region close to the AF transition (i.e., for $T< T_{G}$, the so-called Ginzburg temperature), indicating a D$\Gamma $A critical
exponent $\gamma $ definitely larger than $1$. The non-perturbative nature
of D$\Gamma $A also allows for a treatment of the critical behavior, e.g. the
size of the critical region, as a function of $U$: From our
data it emerges that, in the $U$-range studied, the size of the region where
the critical behavior deviates from the MF predictions (here: from
linearity) increases with $U$.
In order to quantify this statement, we have performed D$\Gamma$A calculations at higher $T$ (upper inset of Fig.\ref{AFsusc})
for $U$ up to $1.5$, and fitted the data linearly in the high-T regime. $T_{G}$ has been hence estimated as
the temperature below which the relative deviation of $\chi_{AF}^{-1}$ from the above-mentioned linear fit becomes larger
than $10\%$ (red arrows in the upper inset of Fig.\ \ref{AFsusc}).
By this criterion for $T_{G}$, the size of the
critical region with non-MF behavior, i.e.  $\Delta T_{crit}= T_{G}- T_N$, increases from $\simeq 0.01$ for $U \! = \! 1.0$, to $\simeq 0.02$ for $U\! = \! 1.25$ and $\simeq 0.025$ for $U\! = \! 1.5$, following therefore the dependence determined by the Ginzburg criterion, which implies the inapplicability of the standard
Landau-Ginzburg expansion in the temperature region $\Delta T_{crit} \propto T_N^2$ \cite{landaubook}. For $U<1$ (not shown) the bending of $\chi_{AF}^{-1}$ becomes hardly visible, since in this regime $T_N \sim \mbox{e}^{-\frac{1}{WU}}$
(with $W \propto 1/D$), and therefore the size of the critical region is rather narrow; the linear behavior for
$U\! > \! 1.5$ becomes confined to temperatures even higher than those shown in Fig.\ \ref{AFsusc}.


A more quantitative study of the critical behavior requires also a
precise evaluation of the critical exponent(s). From the behavior of the
spin-susceptibility, one can extract the values of the critical exponent $
\nu $, which controls the divergence of the AF-correlation length $\xi $
(defined as the square root of the inverse mass of the spin-spin propagator
at $\mathbf{q}=\mathbf{Q},\,\omega \!  = \! 0$) when $T\rightarrow T_{N}$. This can
be computed either from the divergence of $\chi
_{AF}$ (i.e., directly from the data shown in
Fig. \ref{AFsusc}),  using the relation $\gamma \! = \! 2\nu $ \cite{note-eta},
or by extracting from $\chi _{AF}$ the value of $\xi $ by fitting its {\bf q}-dependence for different $T$ \cite{note-fitxi}.

\begin{figure}[tb]
\begin{center}
\includegraphics[width=7.7cm]{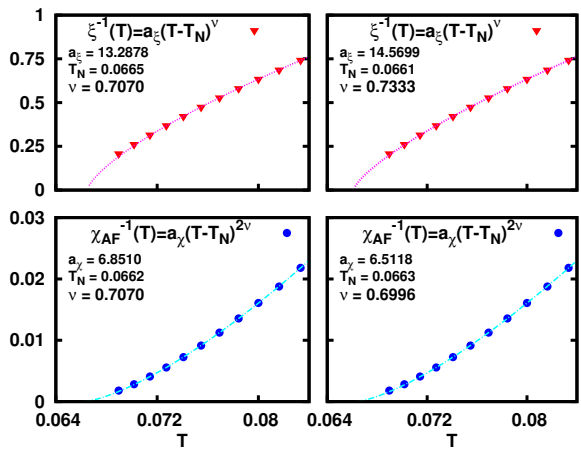}
\end{center}
\caption{(Color online) Fit of $\protect\chi^{-1}_{AF}(T)$ and $\protect\xi^{-1}$(T) for
the highest interaction value considered, i.e., $U=2.5$. Left: fit with fixed $%
\protect\nu=0.707$ (Heisenberg-exponent in $d\! =\! 3$ \cite{collinsbook}). Right: free fit, showing the
good compatibility with the $d\! = \! 3$ Heisenberg universality class.}
\label{fit_exp}
\end{figure}

The results of our analysis, shown in Fig.\ \ref{fit_exp}, demonstrate
that D$\Gamma$A  can describe well the AF criticality of the
Hubbard model. For the largest values of $U=2.5$,
indeed, both divergences of $\chi_{AF}$ and $\xi$ observed in D$\Gamma$A can
be described (left panels of Fig. \ref{fit_exp}) with high-accuracy by the
 critical exponent $\nu=0.707$ of the $d\!=\! 3$-Heisenberg AF.
This is expected to be the correct exponent, not only because the
half-filled Hubbard can be mapped onto the Heisenberg model but also since
dimension and symmetry of the order parameter suggest the same universality
class. Similar results, though with a lower degree of precision, can be
found by directly fitting the value of the $\nu$ exponent to $\chi_{AF}^{-1}$
and $\xi$ (right panels): For $U=2.5$, our two fits provide an estimate of $\nu$  $\sim 0.70$ and $0.73$, respectively. This shows the Heisenberg universality is still  valid also in a parameter region (i.e., at intermediate coupling), where the Hubbard model is not well approximated by the Heisenberg model \cite{Limelette03}.

\begin{figure}[tb]
\begin{center}
\includegraphics[width=8.8cm]{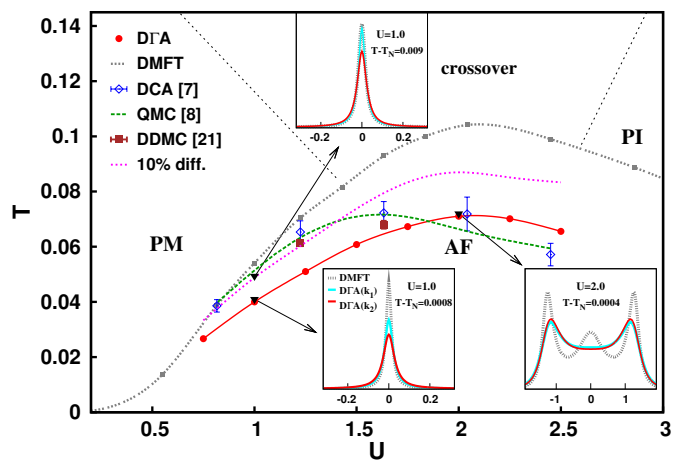}
\end{center}
\par
\vspace{-.5cm}
\caption{(Color online) N\'eel Temperature in D$\Gamma$A, compared with the corresponding DMFT, DCA, QMC and DDMC ones. Also shown is the region where
corrections to DMFT in form of non-local correlations become dominant
 (violet dashed-dotted line: $10$ \% relative change between the DMFT and the D$\Gamma$A self-energies at the lowest Matsubara frequency). Inset: DMFT vs. D$\Gamma$A spectral functions (see text). }
\label{phase-diag}
\end{figure}

A natural by-product of the calculations of the critical exponents is the
determination of $T_N$ at the D$\Gamma$A level, whose values
overall well agree with the most accurate DCA and QMC/DDMC data (see Fig.\ \ref{phase-diag}). The deviations around $U=1$ might originate from neglecting the rather small non-local corrections of the charge- and particle-particle-channels, which could affect non-universal quantities such as $T_N$. On the other hand, also the DCA/QMC finite-size extrapolation is difficult in this region since the  AF correlation length is large.
Let us also note that in this regime the D$\Gamma$A self-energy compares well with the DCA one of Ref.\ \cite{gull_vergl} ($U\! =\! 1.633,T\! =\! 0.0714$): The deviation $\frac{1}{N}\sum_{n} |\mbox{Im}\,\Sigma_{D\Gamma A}({\bf k},\omega_n) -\mbox{Im}\,\Sigma_{DCA}({\bf k},\omega_n)|/|\mbox{Im} \,\Sigma_{D\Gamma A}({\bf k},\omega_n)|$ is $<5\%$  in the sum over the first $N\! = \!7$ Matsubara frequencies (i.e, for those, where a deviation from  DMFT is observable). This is within the DCA difference between the two largest clusters considered ($84$ and $100$ sites).


Finally, we investigate the effects of the non-local corrections on the
spectral properties of the $d\!=\!3$ Hubbard model. On general
grounds, the maximum impact of non-local corrections is to be expected close
to the second-order transition line.
This is because the corresponding spin susceptibility, which
explicitly enters in the D$\Gamma$A equations for $\Sigma$, is
diverging at the transition (red line in Fig.\ \ref{phase-diag}).
Such behavior is particularly evident in the spectra shown in the two lower insets of Fig.\  \ref{phase-diag} for temperatures slightly above the $T_N$ of D$\Gamma$A. Specifically, we compared paramagnetic DMFT and D$\Gamma$A spectral functions at two different ${\bf k}$-points on the Fermi Surface (FS) \cite{noteFS}, i.e. $k_{1}=(\frac{\pi}{2},\frac{\pi}{2},\frac{\pi}{2})$,  $k_{2}=(\pi,0,\frac{\pi}{2})$. At weak-coupling ($U\!= \! 1$) we observe a strong broadening of the DMFT quasiparticle (QP) peak. At $U\! = \! 2$, the enhanced scattering by non-local spin fluctuations  even qualitatively changes the spectra:  the (already) damped QP peak of DMFT is transformed into a ``pseudogap'' in D$\Gamma$A. In principle, one can expect pseudogap behavior very close to the N\'eel temperature also for an arbitrarily small Coulomb interaction.
The corresponding region appears, however, at small $U$ very narrow: a qualitative estimate according to equ. (\ref{SE0})
yields the condition for the pseudogap behavior $\xi >4\pi v_{F} ^{3}/(T_{N}U^{2})$ ($v_{F}$ is an average Fermi velocity),
which can be hardly fulfilled at small $U$, where $T_{N}$ is exponentially small. Outside the pseudogap region,
AF fluctuations yield only an increase of the scattering rate $\gamma({\bf k})\!=\!-\mbox{Im}\, \Sigma({\bf k},\omega\!=\!0)$: e.g., at $T\! \sim \! T_N$ and $U=1$ we obtain $\gamma_{DMFT}=0.02$,  $\gamma_{D\Gamma A}(k_1)=0.033$, and  $\gamma_{D\Gamma A}(k_2)=0.041$.

By increasing $T$, non-local corrections become naturally weaker, since AF-fluctuations are reduced in intensity and spatial extension, see, e.g., the temperature behavior of $\xi$ in Fig. \ref{fit_exp}. As a criterion to evaluate the impact of non-local correlations, valid  for the ``pseudogap'' as well as for insulating spectra, we have chosen  the  relative change between the D$\Gamma$A  and DMFT self-energy  at the lowest Matsubara frequency:
 $|\Sigma_{\rm DMFT}(i\nu_1)-\Sigma_{\rm DGA}({\mathbf k_2},i\nu_1)|/|\Sigma_{\rm DMFT}(i\nu_1)|$. Note that this criterion is directly related to the QP weight
$Z$ in the metallic phase if the linear low frequency behavior of the
self-energy already holds (approximately) at the lowest Matsubara frequency $i\nu_1$.
By this one-particle criterion, DMFT is reliable down to the violet line in  Fig.\ \ref{phase-diag} below which deviations exceed  $10$\%.
Above this line, the impact of the non-local correlations on the spectral functions appears indeed moderate (upper inset of Fig.\ \ref{phase-diag}): this is also confirmed by the analysis of the spectral function, where the QP weight $Z$ is unchanged (within errors) from the DMFT value ($Z=0.76$) and  the enhancement of $\gamma$ is  much smaller than before ($\gamma_{DMFT}=0.027, \, \gamma_{D\Gamma A}(k_1)=0.028, \, \gamma_{D\Gamma A}(k_2)=0.036$).

While our  findings may validate (a posteriori) the usage of DMFT for computing spectral functions in $d\! = \! 3$, provided one is
not interested in the immediate vicinity of (second-order) magnetic instabilities, it is important to note that  the
width of the critical region is not small at intermediate $U$. For instance, we observe that the size of the critical region $\Delta T_{crit}$  at $U\! > \! 1.25$ exceeds the violet line.  Significant effects of non-local correlations may occur even further away from the AF-transition, depending on the quantity
under consideration. In particular relevant deviations from the DMFT predictions at even higher-$T$s have been reported when analyzing the
temperature dependence of the entropy \cite{gull3d_coldat}.

In conclusion, we have analyzed non-perturbatively the effect of non-local
correlations in the $d=3$ half-filled Hubbard model by
means of D$\Gamma$A.
When considering regions where spatial correlations strongly modify the DMFT  physics, which is particularly true close to magnetic instabilities,  D$\Gamma$A represents a very powerful tool for studying the
critical properties beyond the MF/DMFT level:  critical exponents of the Hubbard model are found to be -within the error bars- identical to those of the
$d\! = \! 3$ Heisenberg model, and D$\Gamma$A provides also for a  proper reduction of
$T_N$ w.r.t. the DMFT prediction.
Moreover, since the D$\Gamma$A scheme includes both spatial and temporal electronic correlations in a non-perturbative way, it looks naturally very promising also for future analysis of quantum phase transitions beyond the weak-coupling regime.

We thank E. Gull and S. Fuchs for discussions and exchanging data. We acknowledge financial support from the EU-India network
MONAMI (GR),  Austria-Russia FWF project I 610-N16 (AT),
RFBR grants no.\ 10-02-91003-ANF\_a, 11-02-00937\--a and Max-Planck associated partner group (AK), and FWF SFB ViCoM F41 (KH). Calculations were done on the Vienna Scientific Cluster.




\begin{thebibliography}{99}

\bibitem{Hubbard63a} J.~Hubbard, \newblock Proc. Roy. Soc. London A \textbf{
276} 238 (1963); M.~C. Gutzwiller, \newblock Phys. Rev. Lett. \textbf{
10} 159 (1963);  J.\ Kanamori, Progr.\ Theor.\ Phy.\ \textbf{30}, 275
(1963).

\bibitem{DMFT} W.Metzner and D.Vollhardt, \newblock Phys. Rev. Lett. \textbf{%
62}, 324 (1989).

\bibitem{DMFT2} A. Georges and G. Kotliar, \newblock Phys. Rev. B \textbf{45}%
, 6479 (1992).

\bibitem{DMFTreview} A. Georges {\sl et al.},
Rev. Mod. Phys. \textbf{68}, 13 (1996).

\bibitem{kinks} K.~Byczuk, {\sl et al.}, \newblock Nature Physics \textbf{3}, 168 (2007); A. Toschi, {\sl et al.} \newblock Phys. Rev. Lett., {\bf 102}, 076402 (2009).

\bibitem{Maier04} T.~Maier  {\sl et al.},\ Rev. Mod. Phys. \textbf{77} 1027 (2005); G.~Kotliar  {\sl et al.}, Phys. Rev. Lett. {\bf 87} 186401 (2001);
A.~I. Lichtenstein and M.~I. Katsnelson, Phys. Rev. B {\bf 62} 9283 (R) (2000).

\bibitem{Kent05a} P. R. C. Kent {\sl et al.},
Phys. Rev. B \textbf{72}, 060411 (2005).


\bibitem{Staudt00} R. Staudt, M. Dzierzawa, and A. Muramatsu, Eur. Phys. J.
B \textbf{17}, 411 (2000).


\bibitem{sigma_k}  E.~Z. Kuchinskii, I.~A. Nekrasov and M.~V. Sadovskii, \newblock Sov. Phys. JETP Lett. {\bf 82} 98 (2005).

\bibitem{DGA1} A.~Toschi, A.~A. Katanin, and K.~Held. {Phys. Rev. B} \textbf{
75}, 045118 (2007); {Prog. Theor. Phys. Supp.} \textbf{176}, 117 (2008).

\bibitem{Kusunose06a} H.~Kusunose, {J. Phys. Soc. Jpn.} \textbf{75}, 054713
(2006).

\bibitem{Slezak06a} C.~Slezak {\sl et al.}
J. Phys.: Condens. Matter {\bf 21}, 435604 (2009).

\bibitem{DualFermion} A. N. Rubtsov, M. I. Katsnelson, and A. I.
Lichtenstein, Phys. Rev. B \textbf{77}, 033101 (2008).

\bibitem{DGA2} A.\ A. Katanin, {A. Toschi}, and K. Held, Phys. Rev. B
\textbf{80}, 075104 (2009).

\bibitem{DualFermion2} S. Brener {\sl et al.}, Phys. Rev. B \textbf{77}, 195105 (2008).

\bibitem{notaDOS} With this choice, for $D=1$, the standard deviation of our $d=3$-DOS is equal to 0.5
as in several previous DMFT and D$\Gamma$A calculations.

\bibitem{landaubook} L.~D.~Landau and E.~Lifschitz, ``Statistical Physics'',
  volume V, p. 476, Eq. (146.15), Pergamon Press (1980).








\bibitem{note-eta} Note that within Moriyasque D$\Gamma$A, the index $\eta$ is not changed from its MF value (i.e., $0$), since the explicitly $\mathbf{q}$%
-dependent terms of the spin-spin propagator (but not its mass!) is computed
at the level of DMFT.


\bibitem{note-fitxi} The value of $\xi$ has been computed by fitting the
D$\Gamma$A spin susceptibility $\chi(\mathbf{q},\Omega=0)$ with
the fitting function $\chi_{fit}=A/[4(\sin^2(\frac{q_x-\pi}{2})+\sin^2(\frac{q_y
-\pi}{2})+\sin^2(\frac{q_z-\pi }{2})) + \xi^{-2}]$.


\bibitem{collinsbook} M.F. Collins, ``Magnetic Critical Scattering'', Oxford
  University Press, New York, 1989.

\bibitem{Limelette03}
Note: For the Mott-Hubbard transition in paramagnetic Cr-doped V$_2$O$_3$ deviations from MF critical exponents have been found only in a tiny parameter region , see P.~Limelette {\sl et al.},
\newblock Science {\bf 302}, 89 (2003).


\bibitem{gull3d_coldat} S. Fuchs et al., Phys. Rev. Lett. {\bf 106}, 030401 (2011).

\bibitem{gull_vergl}  E. Gull {\sl et al.}, Phys. Rev. B {\bf 83}, 075122 (2011).

 \bibitem{noteFS} This choice is highly significant, because for these $\mathbf{k}$-points (at the FS) the
 largest/smallest deviations from the DMFT self-energy are found.

\bibitem{gull2d} E. Gull {\sl et al.}, Phys. Rev. B {\bf 82}, 155101 (2010).





\end{thebibliography}
\end{document}